\pdfoutput=1
\documentclass[twoside,leqno,twocolumn]{article}
\usepackage{amsmath,bm}
\usepackage{amsfonts}
\usepackage{latexsym}
\usepackage{array}
\usepackage{graphicx}
\usepackage{url}
\usepackage{verbatim}
\usepackage{lscape}
\usepackage{rotating}
\usepackage{longtable} 

\begin{document}

\title{MeSH descriptors indicate the knowledge growth in the SARS-CoV-2/COVID-19 pandemic.}
\author{Johannes Stegmann\footnote{Member of the Ernst-Reuter-Gesellschaft der Freunde, F\"orderer und Ehemaligen  der Freien Universit\"at Berlin e.V.}
\footnote{Former (now retired) employee of the Medical Library of the Free University Berlin and the Charit\'e Berlin.}
\footnote{Radebeul, Germany, johannes.stegmann@fu-berlin.de} }
\date{}
\maketitle

\begin{abstract} 
The scientific papers dealing with the novel betacoronavirus SARS-CoV-2 and the coronavirus disease 2019 (COVID-19) caused by this virus, published in 2020 and recorded in the database PUBMED, were retrieved on April 27, 2020. About 20\% of the records contain Medical Subject Headings (MeSH), keywords assigned to records in the course of the indexing process in order to summarise the articles' contents. The temporal sequence of the first occurrences of the keywords was determined, thus giving insight into the growth of the knowledge base of the pandemic. 
\end{abstract}
\newline 

\section{Introduction} \renewcommand*{\thefootnote}{\fnsymbol{footnote}}
\indent 
\indent The rapid worldwide spread of the new epidemic COVID-19, caused by the virus SARS-CoV-2, with now more than 3.4 million confirmed cases of the disease and a confirmed death rate of almost 7\% (World Health Organization, 2020) requires fast and comprehensive efforts of states and societies to combat the disease effectively by means of practical and appropriate medical, administrative and economic actions. Moreover, the scientific community has the responsibility to bundle resources and manpower to develop tests, drugs and vaccines in order to gain control over the virus and the disease as quick as possible. \\
An important research tool is immediate and unlimited access to the scientific literature. For the biomedical specialties, the freely available database PUBMED/MEDLINE\footnote[1]{provided by the U.S. National Center for Biotechnology Information, NCBI, www.ncbi.nlm.nih.gov/pubmed/} is indispensable for a comprehensive retrieval of the published scientific papers on biomedical research questions. Besides the bibliographic metadata (as author name(s), publication year, journal name and volume, etc.) PUBMED records are indexed by a controlled vocabulary of many thousands descriptors, the Medical Subject Headings (MeSH\footnote[2]{PUBMED's hierarchichal thesaurus, the U.S. National Library of Medicine's controlled vocabulary, www.ncbi.nlm.nih.gov/mesh}). In addition to the words contained in titles and abstracts of indexed papers, the MeSH descriptors assigned to PUBMED records are significant for a thorough analysis of the papers' content.\\
In the study presented here the publications on SARS-CoV-2 and COVID-19 were retrieved and downloaded from PUBMED. The MeSH descriptors were extracted from the records already annotated with MeSH. The keywords were ordered chronologically according to the publication date of their associated papers and their first occurences were determined.

\begin{table*}[hbtp]
\caption{SARS-CoV-2/COVID-19 papers without and with MeSH terms retrieved\textsuperscript{*} from PUBMED.}
*\footnotesize Date of retrieval: April 27, 2020
\centering
\begin{tabular}{rcc}
\noalign{\smallskip}
\hline
\noalign{\smallskip}
\hspace{5.0cm} All papers & Papers with MeSH terms  & Number of distinct MeSH terms \\
\noalign{\smallskip} 
\noalign{\smallskip}\hline\noalign{\smallskip}
7366 & 1504 & 1769  \\
\noalign{\smallskip} 
\noalign{\smallskip}\hline
\end{tabular}
\end{table*} 

\section{Methods} \renewcommand*{\thefootnote}{\fnsymbol{footnote}}
\subsection{Retrieval} \indent
\indent Papers published 2020 were retrieved and downloaded from PUBMED on April 27, 2020, using the following search profile:\\
{\em new coronavirus* OR novel coronavirus* OR ncov OR sars-cov OR covid* OR cov-2 OR cov-19} (the truncation asterisk - "*" - retrieves all terms with that word stem).

\subsection{Information extraction} \indent
\indent The Medical Subject Headings (MeSH) assigned to PUBMED records are contained in the MH fields. Controlled vocabulary is also contained in the RN fields. The contents of both fields were extracted from the records. In addition, the unique record numbers and the database indexing date were extracted from the PMID and MHDA fields, respectively.

\subsection{Programming, calculations} \indent
\indent Extraction of record field contents, clustering, data analysis, calculations and visualisation were done using homemade programs and scripts for perl (version 5.26.1) and the software package R version 3.4.4 (R Core Team, 2018). All operations were done on a commercial PC run under Ubuntu version 18.04 LTS.

\begin{figure}[htpb]
\centerline{
\includegraphics[height=8.0cm]{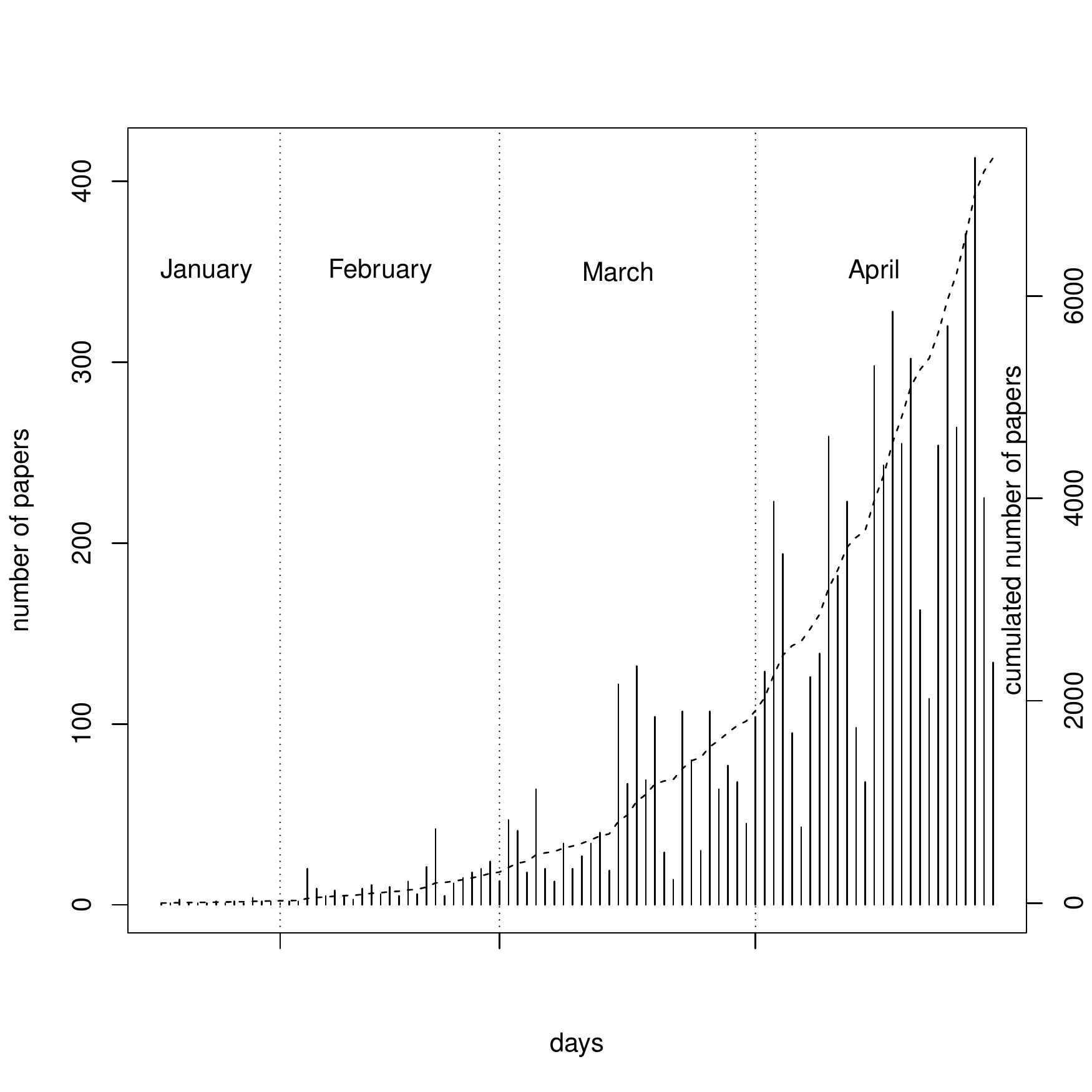}}
\caption{Number of SARS-CoV-2/COVID-19 papers indexed daily\textsuperscript{*} in PUBMED.}
  *\footnotesize January to April-27, 2020. Days without papers are omitted. \newline
dashed line: cumulated number of papers.
\label{fig:fig01}
\end{figure}

\begin{figure}[htpb]
\centerline{
\includegraphics[height=8.0cm]{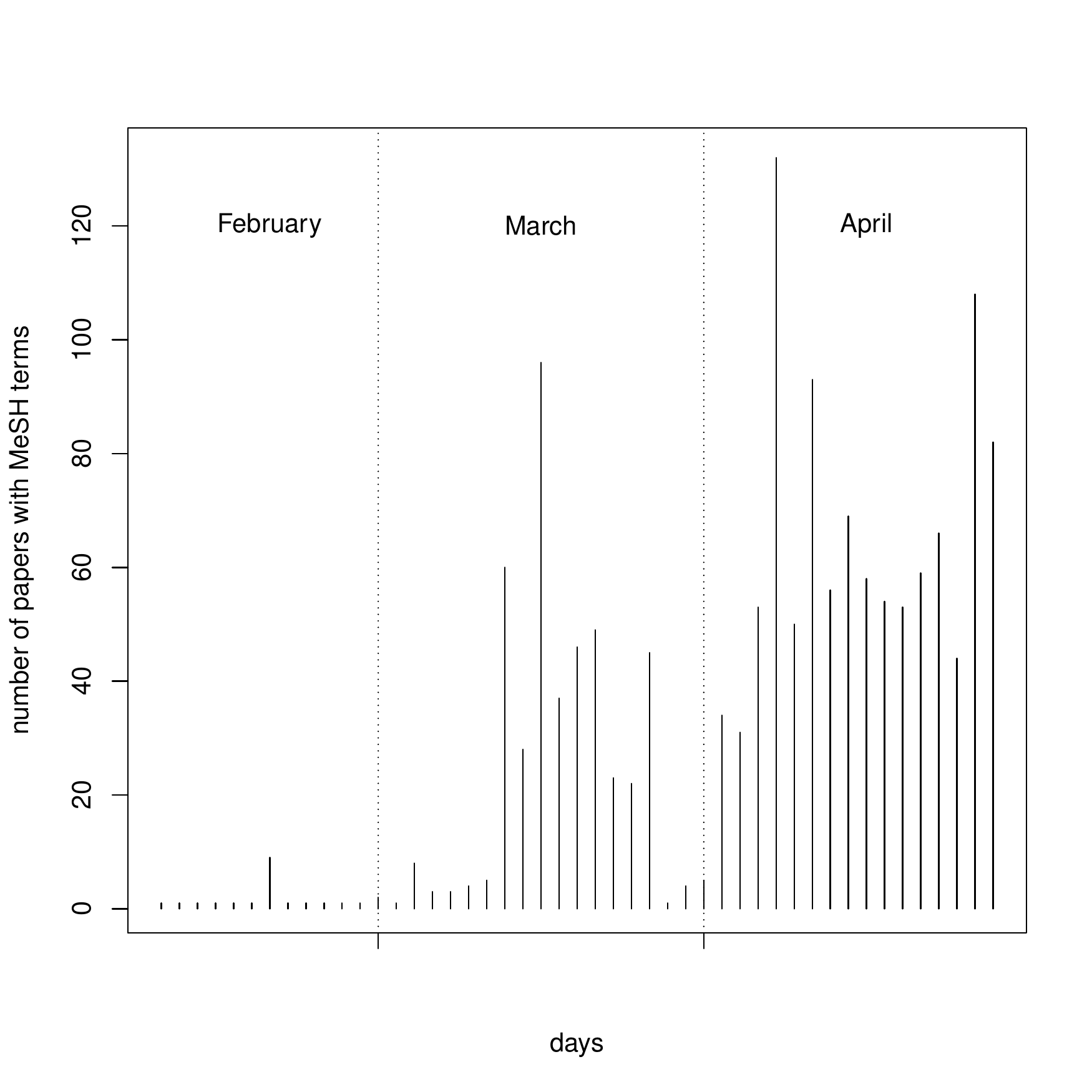}}
\caption{Daily\textsuperscript{*} number of SARS-CoV-2/COVID-19 papers indexed with MeSH terms in PUBMED.}
  *\footnotesize February to April-27, 2020. Days without papers are omitted.
\label{fig:fig02}
\end{figure}

\begin{figure}[htpb]
\centerline{
\includegraphics[height=8.0cm]{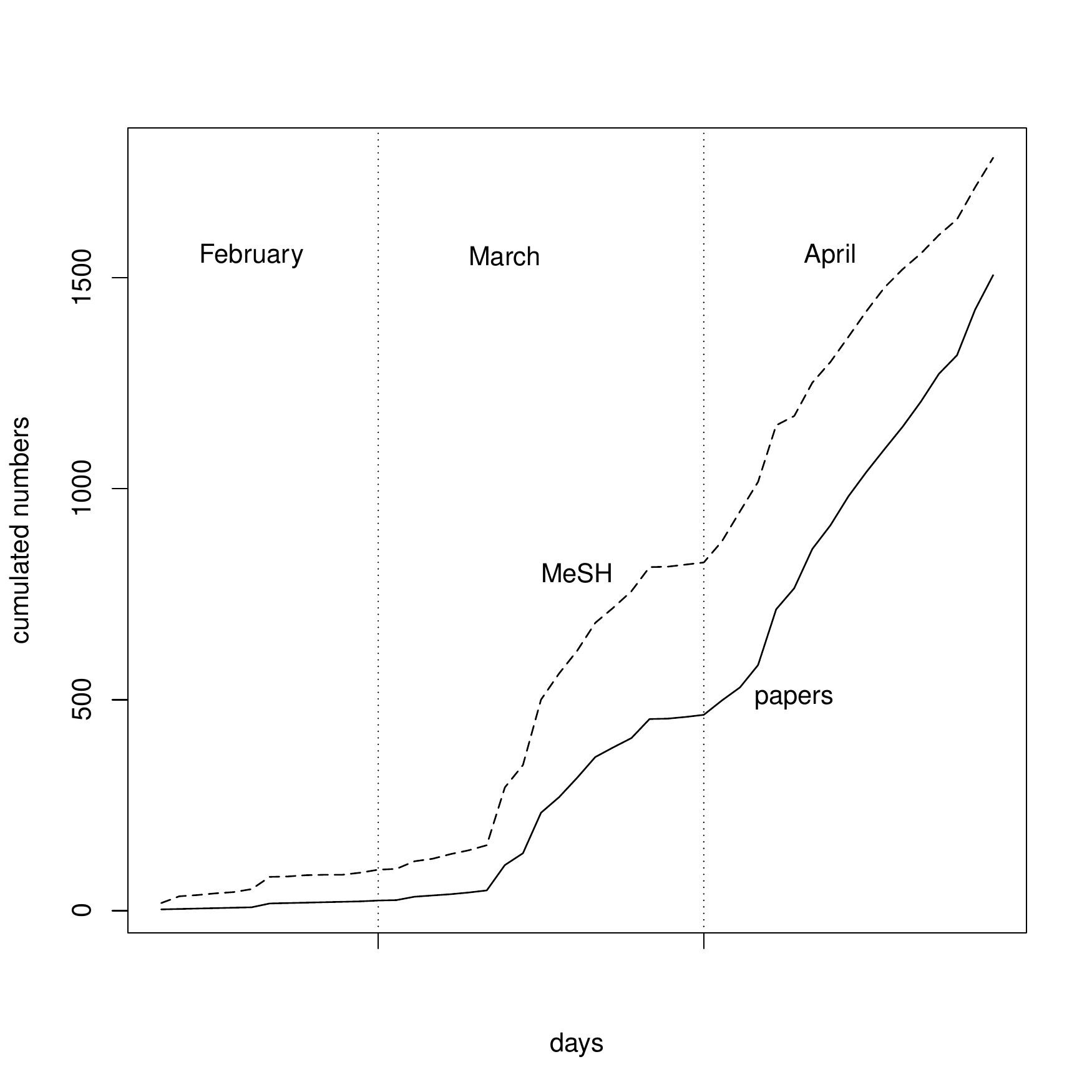}}
\caption{Cumulated daily\textsuperscript{*} number of SARS-CoV-2/COVID-19 papers with assigned MeSH terms and their distinct new MeSH terms.}
  *\footnotesize February to April-27, 2020.
\label{fig:fig03}
\end{figure}

\section{Results and Discussion} 

The search profile mentioned in the Methods paragraph retrieved 7366 publications for the period January to April 27, 2020 (Table 1). The daily distribution of the items is shown in Figure 1. In January and February 2020 few papers appeared, followed by a considerable publication boost in March and April. Similar data were published by Kousha and Thelwall (2020) and Torres-Salinas (2020). About 20\% of the papers have already assigned MeSH terms (Table 1). The daily distribution of papers with MeSH terms (Figure 2) is more or less similar to that of all papers (compare Figure 2 and Figure 1). Figure 3 shows that the cumulation of MeSH terms parallels the cumulation of MeSH papers. Both, Figure 2 and Figure 3, indicate that the MeSH terms used to classify SARS-CoV-2/COVID-19 papers may exhibit to some extent the knowledge accumulation and development around the pandemic. 

\begin{table*}[htpb]
\caption{Temporal increase of distinct MeSH terms assigned to SARS-CoV-2/COVID-19 papers.}
\centering
\begin{tabular}{ccccc|ccccc}
\noalign{\smallskip}
\hline
\noalign{\smallskip}
Date & \multicolumn{2}{c}{Papers with MeSH} & \multicolumn{2}{c}{New MeSH terms} & Date & \multicolumn{2}{c}{Papers with MeSH} & \multicolumn{2}{c}{New MeSH terms}\\
dd mm yyyy & number & cumulated  & number  & cumulated & dd mm yyyy & number  & cumulated   & number   & cumulated  \\
\noalign{\smallskip} 
\noalign{\smallskip}\hline\noalign{\smallskip}
06 02 2020 & 1 & 1  &  3 &  3 & 21 03 2020 & 46 & 313 & 54 & 601  \\
08 02 2020 & 1 & 2  & 16 & 19 & 24 03 2020 & 49 & 362 & 66 & 667  \\
11 02 2020 & 1 & 3  &  3 & 22 & 25 03 2020 & 23 & 385 & 36 & 703    \\
14 02 2020 & 1 & 4  &  4 & 26 & 27 03 2020 & 22 & 407 & 39 & 742    \\
18 02 2020 & 1 & 5  &  3 & 29 & 28 03 2020 & 45 & 452 & 57 & 799   \\
19 02 2020 & 1 & 6  &  7 & 36 & 28 03 2020 & 1 & 453 & 1 & 800   \\
20 02 2020 & 9 & 15 & 29 & 65 & 31 03 2020 & 4 & 457 & 5 & 805    \\
23 02 2020 & 1 & 16 & 1 & 66 & 01 04 2020 & 5 & 462  & 5 & 810  \\
25 02 2020 & 1 & 17 & 3 & 69 & 02 04 2020 & 34 & 496 & 50 & 860   \\
26 02 2020 & 1 & 18 & 1 & 70 & 03 04 2020 & 31 & 527 & 71 & 931   \\
27 02 2020 & 1 & 19 & 0 & 70 & 04 04 2020 & 53 & 580 & 70 & 1001    \\
29 02 2020 & 1 & 20 & 5 & 75 & 09 04 2020 & 132 & 712 & 134 & 1135    \\
03 03 2020 & 2 & 22 & 7 & 82& 10 04 2020 & 50 & 762 &  22 & 1157    \\
04 03 2020 & 1 & 23 & 2 & 84 & 11 04 2020 & 93 & 855 & 79 & 1236   \\
07 03 2020 & 8 & 31 & 18 & 102 & 14 04 2020 & 56 & 911 & 49 & 1285   \\
10 03 2020 & 3 & 34 &  6 & 108 & 15 04 2020 & 69 & 980 & 60 & 1345   \\
11 03 2020 & 3 & 37 & 11 & 119 & 16 04 2020 & 58 & 1038 & 61 & 1406   \\
13 03 2020 & 4 & 41 & 9 & 128 & 17 04 2020 & 54 & 1092 & 56 & 1462    \\
14 03 2020 & 5 & 46 & 12 & 140 & 18 04 2020 & 53 & 1145 & 43 & 1505    \\
17 03 2020 & 60 & 106 & 137 & 277 & 21 04 2020 & 59 & 1204 & 37 & 1542    \\
18 03 2020 & 28 & 134 & 53 & 330 & 22 04 2020 & 66 & 1270 & 44 & 1586   \\
19 03 2020 & 96 & 230 & 155 & 485 & 23 04 2020 & 44 & 1314  & 37 & 1623    \\
20 03 2020 & 37 & 267 & 62 & 547 & 24 04 2020 & 108 & 1422  & 76 & 1699    \\
           &    &    &     &     & 25 04 2020 & 82 & 1504 &  70 & 1769    \\

\noalign{\smallskip} 
\noalign{\smallskip}\hline
\end{tabular}
\end{table*}

\begin{table*}[htpb]\small
\caption{SARS-CoV-2/COVID-19 papers 2020: Temporal sequence of MeSH terms appearance (examples).}
\centering

\begin{tabular}{cp{5cm}|cp{9cm}}
\noalign{\smallskip}
\hline
\noalign{\smallskip}
Date & MeSH terms & Date & MeSH terms \\
\noalign{\smallskip} 
\noalign{\smallskip}\hline\noalign{\smallskip}
February 06 & INFORMATION DISSEMINATION & March 17 & LOPINAVIR-RITONAVIR DRUG COMBINATION \\ 
            & DISEASE OUTBREAKS         &          & RITONAVIR \\
            & CORONAVIRUS INFECTIONS    &          & LOPINAVIR \\
February 08 & BETACORONAVIRUS           &          & CHLOROQUINE \\
            & COVID-19                  &          & REMDESIVIR \\  
            & SEVERE ACUTE RESPIRATORY SYNDROME CORONAVIRUS 2 &   & FECES  \\      
            & PANDEMICS                 &          & COVID-19 SEROTHERAPY \\
February 14 & EMERGENCIES               &          & SPIKE GLYCOPROTEIN, CORONAVIRUS \\
            & CHINA                     &          & SARS-COV-2  \\     
            & GLOBAL HEALTH             & March 18 & ASYMPTOMATIC INFECTIONS \\
February 19 & CONTACT TRACING           &          & INFECTIOUS DISEASE TRANSMISSION, PATIENT-TO-PROFESSIONAL \\
            & TRAVEL-RELATED ILLNESS    &          & CLUSTERED REGULARLY INTERSPACED SHORT PALINDROMIC REPEATS \\
February 20 & COVID-19 DIAGNOSTIC TESTING & March 19 & MOUTH MUCOSA \\         
            & COVID-19 DRUG TREATMENT     & March 20 & VULNERABLE POPULATIONS \\         
            & MULTIPLE ORGAN FAILURE      &          & HOMES FOR THE AGED \\      
February 29 & TOMOGRAPHY, X-RAY COMPUTED  & March 27 & MOLECULAR DOCKING SIMULATION \\
            & RADIOGRAPHY, THORACIC       & April 14 & INFECTIOUS DISEASE TRANSMISSION, PROFESSIONAL-TO-PATIENT \\         
March 10    & UNITED KINGDOM              &          & WASTEWATER-BASED EPIDEMIOLOGICAL MONITORING \\
            & ITALY                       & April 16 & MESENCHYMAL STEM CELLS \\
March 14    & AGED, 80 AND OVER           &          & FECAL MICROBIOTA TRANSPLANTATION \\
            & AGE DISTRIBUTION            & April 21 & OLFACTION DISORDERS \\        
March 17    & QUARANTINE                  & April 24 & TASTE DISORDERS \\
            & HOSPITAL BED CAPACITY       &          & BLOOD COAGULATION DISORDERS \\
            & ZOONOSES                    & April 25 & CRISPR-CAS SYSTEMS \\       

\noalign{\smallskip} 
\noalign{\smallskip}\hline
\end{tabular}
\end{table*} 

\subsection{Temporal development of MeSH descriptors} \indent
\indent
Tables 2 and 3 show the addition of (new) MeSH terms to indexed papers by day. Table 2 shows the numbers, Table 3 examples of the terms. Table 2 lists the dates and the number of publications with MeSH indexed as well as the numbers of new MeSH terms, i.e. MeSH terms which are not contained in the papers of the preceding dates. In Table 3 selected Medical Subject Headings are listed according to the sequence of their appearance from February to April 2020. The MeSH terms assigned to papers in the first half of February 2020 indicate the knowledge of a disease outbreak in China of pandemic proportions, caused by a betacoronavirus, and both, disease and virus, are already labelled (Table 3). The concomitant - possibly life-threatening - implications of the new disease, disease-spreading mechanisms, necessary diagnostic tools, assessment of especially vulnerable age groups, problems of health care systems, possible drug therapy schemes and other therapy approaches become evident using the information contained in MeSH terms assigned to papers published in subsequent days, weeks and months. Although the fraction of papers with assigned MeSH terms is relatively low (see Table 1), may the whole set of already more than 1700 MeSH terms (at the download date, see e.g. Table 2) greatly benefit (not only) medical experts.

\section{Conclusion} \indent
\indent The short analysis of SARS-CoV-2/COVID-19 publications presented here shows that careful inspection of the assigned Medical Subject Headings is worthwhile and associated with an increase of the knowledge base of the pandemic.


\end{document}